# A COMPARISON OF APPROACHES FOR TRAFFIC ENGINEERING IN IP AND MPLS NETWORKS

Alan Gous, Arash Afrakhteh, Cariden Technologies
John Evans, Cisco

## ABSTRACT

As IP / MPLS network service providers drive for improved network efficiencies, traffic engineering is becoming increasingly important. In this article we review the approaches for traffic engineering in IP / MPLS networks and present the result of a study, which compares their performance in operational networks.

## 1. INTRODUCTION

In conventional IP networks Interior Gateway routing Protocols (IGPs) such as OSPF [RFC 2328] and IS-IS [RFC1142] forward IP packets on the shortest cost path towards the destination IP subnet address of each IP packet. The computation of the shortest cost path is based upon a simple additive metric (also known as weight or cost), where each link has an applied metric, and the cost for a path is the sum of the link metrics on the path. In many networks the metrics are set according to the physical link capacity or link delay; in these cases actual available (i.e. unutilised) bandwidth is not taken into account and consequently, traffic can aggregate on the shortest (i.e. lowest cost) path, potentially causing links on the shortest path to be congested while links on alternative paths are under-utilised.

Consider, for example, the network in Figure 1, where each link is 10Gbps and each link has the same metric (assume a metric of 1). If there were a traffic demand of 5Gbps from R1 to R8, and a traffic demand of 7Gbps from R2 to R8, then the IGP would pick the same route for both traffic demands, i.e. R1/R2→R3→R4→R7→R8, because it has a metric of 4 (summing the metrics for each of the links traversed) and hence is the shortest path.

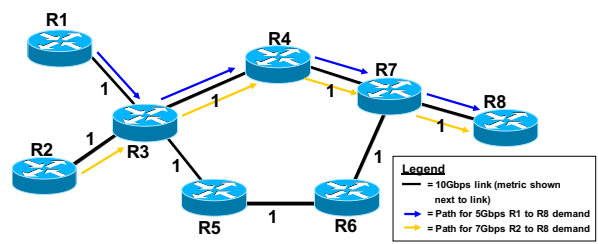

**Figure 1.** Traffic Engineering: the problem.

It is often said that network planning is the process of fitting your network to your traffic, while traffic engineering (TE) is the process of fitting your traffic to your network. Given a fixed network topology, a traffic engineer will set, or influence, the routes taken by the traffic through that network, to make best use of the capacity available in that topology. In practice, this involves balancing a number of specific quality of service criteria as traffic patterns change and grow, including latency and utilisation levels, as well as resiliency and speed of recovery under failure. In IP networks today, traffic engineering is often considered synonymous with Multi-Protocol Label Switching (MPLS) or more specifically, with MPLS Traffic Engineering (MPLS-TE). IP routing, however, can also be engineered without requiring MPLS, primarily through the manipulation of IGP metrics in the network.

The primary benefit of traffic engineering is one of cost saving, i.e. in terms of requiring less capacity to achieve the same service levels when compared to the non-TE case, or in more aggregate traffic being supported for the same provisioned bandwidth as the non-TE case. Network costs also include operational and management costs, however, so any potential bandwidth savings need to be balanced against the additional cost and complexity of deploying a particular traffic engineering solution.

In this article, we provide an overview of the different approaches for traffic engineering in IP / MPLS networks, describe an approach for comparing their efficacy and present the results of a comparative study conducted on a number of operational IP networks. We define the goals for traffic engineering and assess how the different traffic engineering approaches compare, with respect to one particularly important criterion: achievable utilisation levels, both under normal operation and failure operation.

## 2. IGP Metric-based Traffic Engineering

The tactical and ad hoc tweaking of IGP metrics to change the routing of traffic and relieve congested hotspots has long been practised in IP networks. Most IP network operations engage in some kind of IGP metric-based traffic engineering through manual manipulation of link metrics.

One trial-and-error approach is to raise metrics, little by little, on links with high utilisations, until the utilisations drop to acceptable levels. Another approach is to construct simple Equal Cost Multipath (ECMP) IGP routes between various pairs of nodes in the network, to load-balance the traffic between them. By setting metrics so that two or more paths between a pair of nodes are of equal length (i.e., the sum of the link metrics are equal), IGP routing will split the traffic equally between these paths. This is a simple procedure to implement if the paths are just one or two circuits in length, but rapidly becomes unwieldy for paths with larger numbers of links, when the effect of the metric changes on all other traffic using these links must also be assessed.

Hence, it is clear that such ad hoc IGP metric-based TE approaches have limited applicability. Further, without knowledge of the network traffic demand matrix – the aggregated ingress to egress traffic flows across the network –, no a priori assessment of how proposed metric changes will affect utilisations can be made. Until relatively recently, it has not been easy to measure or determine the network traffic demand matrix. As a consequence of these factors, for a long time, TE based on IGP metric manipulation was not considered viable for systematic network wide traffic engineering and it was often cited that changing the link metrics just moves the problem of congestion around the network.

Any more sophisticated manipulation of metrics must therefore be done with knowledge of the network traffic demand matrix. Demand level measurements of traffic have traditionally not been available in pure IP networks, however, a number of options are now available to obtain this information as described in section 4.4.

In parallel there has been an increase in research in the use of systematic (i.e. network wide) traffic engineering by manipulating IGP metrics. Various optimisation procedures can be used tackle the problem. Variations of optimisation algorithms range from sophisticated local search [1] to attempts at linear programming-(LP)[1] based solutions [3], similar in structure to the multicommodity-flow network routing problems. Further, IGP metric based traffic engineering has been realised in the development of automated planning tools, which take inputs of the network logical (i.e. IGP) and physical topology, together with the network traffic demand matrix and derive a more optimal set of link metrics based upon a defined optimisation goal. These optimisation goals may be to minimise the maximum utilisation on aggregate, or per class, i.e. where the Differentiated Services Architecture [RFC2547] is deployed.

IGP metric-based TE chooses from all possible configurations of metrics in a network to select optimal routings. Opinions vary as to the efficiency of IGP metric-based TE compared to other approaches MPLS-TE. For example, [4] present specially constructed network topologies and demands for which IP routing can be made to do arbitrarily badly compared to explicit routing. On the other hand, evaluating the performance of their local search heuristic, [1] conclude that they can find metric settings that "get within a few percent of the best possible with general routing, including MPLS."

# 3. RSVP-BASED MPLS TRAFFIC ENGINEERING

Unlike IGP metric-based IP routing, which uses pure destination-based forwarding, MPLS-TE makes use of the implicit MPLS characteristic of separation between the data plane (also known as the forwarding plane) and the control plane to allow routing decisions to be made on criteria other than the destination address in the IP packet header, e.g. such as available link bandwidth. MPLS-TE provides constraint based path computation and explicit routing capabilities, which can be used to divert traffic away from congested parts of the network to links where bandwidth is available and hence potentially make more optimal use of available capacity. Label switched paths (LSPs), which are also termed "traffic engineering tunnels" in the context of MPLS-TE, are used to steer traffic through the network allowing links to be used which are not on the IGP shortest path to the destination.

MPLS-TE makes use of the following mechanisms:

1. *Resource / policy information distribution.* Each router within the network floods information on the available bandwidth resources for its connected links, together with administrative policy constraint information, throughout the network by means of extensions to link-state based IGP routing protocols such as OSPF [RFC3630] and IS-IS [RFC3784]. As TE tunnels are unidirectional, each TE-enabled router maintains a pool of available (i.e. currently unused) TE bandwidth in the egress direction for each interface that it has. All of the routers within the MPLS-TE area will receive the information on the available network resources, advertised via OSPF or IS-IS.

---

[1] See [2] for an introduction to LP-based network flow algorithms.

2. *Constraint based path computation.* With MPLS-TE, tunnel paths can be dynamically calculated online in a distributed fashion by the TE tunnel sources (known as tunnel "head-ends") themselves – this is referred to as a *dynamic path option* – or can be determined by an offline centralised function (also known as a tunnel server or path computation element) which then specifies the explicit tunnel path a head-end should use for a particular tunnel – this is referred to as an *explicit path option*. With either approach, a constraint-based routing (CBR) algorithm uses a constraint-based shortest path first (CSPF) calculation to determine the path that a particular tunnel will take based upon a fit between the available network bandwidth resources (and optionally policy constraints) and the required bandwidth (and policies) for that tunnel. This CSPF calculation is similar to a conventional IGP shortest path first (SPF) calculation, but also takes into account bandwidth and administrative constraints, pruning links from the topology if they advertised insufficient resources, i.e. not enough bandwidth for a tunnel, or if they violate tunnel policy constraints. The shortest (i.e. lowest cost) path is then selected from the remaining topology.
3. *RSVP for tunnel signalling.* Whether online or offline path calculation is used, the output is an explicit route object (ERO) which defines the hop-by-hop path the tunnel should take. The ERO is handed over to the Resource ReSerVation Protocol (RSVP) [RFC2205] – with enhancements for MPLS-TE [RFC3209] – in order to signal the tunnel LSP.

Far more possible routing configurations are possible with MPLS-TE, in which a route for each demand can be chosen separately, and independently of the others, that are possible with IGP metric-based TE. So it is reasonable to assume that MPLS-TE should be able to achieve more optimal routings, at least under the no-failure scenario, than IP routing. We note, however, that as illustrated by the results in [5] more optimal solutions are generally achievable with *explicit path option*s than *dynamic path options;* with explicit path options a tunnel server/path computation element is able to calculate the best routings for all tunnels, whereas with dynamic path options each head-end is only able to determine the routing of paths for tunnels for which it is the head-end. In the remainder of this paper we focus on explicit path options.

When failure scenarios are also considered, we may expect MPLS-TE with explicit path options to perform even better than the no-failure scenario. IGP routing is dependent on the same metric settings to choose routings under any failure scenario. TE tunnels with explicit path options allow all individual demands to be routed independently of one another, under failure scenarios in which circuits containing these demands fail. In particular, when using primary and secondary explicit pairs, there is complete freedom in the choice of each secondary route.

On the other hand, managing MPLS-TE explicit paths in an IP network is more complex than pure IP routing: a mesh of MPLS TE tunnels must be provisioned and managed. So the question arises: does the increase in efficiency of utilisation of MPLS-TE with explicit path options justify the additional administrative burden?

## 4. COMPARING APPROACHES

In fact, the difference in performance between these approaches depends hugely on the network topology and traffic demand matrix selected for evaluation. So the most relevant question becomes, how do they compare in actual network topologies? This question can be hard to answer, because details of most currently operating network topologies which are large enough for TE in general to make a substantial impact, are not publicly available. The details of the networks used in the study that follows are not publicly available either, and therefore just the methods used in the study, and the summarized results, may be made public.

## 4.1 Defining The Goals

The problem of traffic engineering can be defined as a mathematical optimization problem; that is a computational problem in which the objective is to find the best of all possible solutions. Given a fixed network topology and a fixed source-to-destination traffic demand matrix to be carried, the optimisation problem could be defined as determining the routing of flows that makes most effective use of capacity. In order to solve this problem, however, it is important to define what is meant by the objective "most effective". In considering the deployment of traffic engineering mechanisms, it is imperative that the optimisation objectives are defined, in order to understand what benefits the different approaches for traffic engineering can provide, and where traffic engineering will not help, but rather more bandwidth is required. There are various criteria that may be used to evaluate different traffic engineering methods. Here we describe one formal benchmark, worst-case utilisation, which will allow the performance of different traffic engineering strategies to be compared.

## 4.2 Utilisation under Failure

Modern IP networks are subject to failure. In large IP / MPLS networks it is unusual to have a week without the failure of some network element, be it an interface card going down, or a fibre cut. These networks need sufficient capacity to deal with the traffic rerouting that will occurs under such failure events.

A useful way to characterize the resilience of a network to failure is to calculate certain *worst-case utilisation* levels under various sets of interesting *failure scenarios*. Each failure scenario is a particular way in which the network may fail. There are three sets of failure scenarios that are commonly taken into account:

1. *Single circuit failures:* a circuit may fail on its own, for example due to a failure in the underlying transport network.
2. *Single node failures:* a node (e.g. router) may fail on its own.
3. *Shared Risk Link Group (SRLG) failure.* Groups of circuits that may be taken down by a single event. For example, a fibre cut will fail all the circuits traversing the fibre. Components that have common failure dependencies are defined as belonging to the same SRLG.

Many more sets of scenarios could be enumerated, such as all pairs of circuits, or site failures for example.

Knowing the network topology, the traffic demand matrix, and the network routing behaviour, it is possible to calculate, through network simulation, the traffic utilisation in each link (or, equivalently, interface) in the network, under any given failure scenario.

## 4.3 Worst-Case Utilisation

Given a particular link, the *worst-case utilisation* under a set of failure scenarios is defined to be the maximum utilisation for that link over all of the failure scenarios in the set. Worst-case link utilisations are useful for identifying bottlenecks in the network that will only become apparent once some failure occurs. The overall network *worst-case utilisation* is a single number which summarizes how resilient the network is, under a given traffic engineering configuration, to network failure. This number is simply the maximum worst-case link utilisation over all links in the part of the network under evaluation. For example, the maximum might be taken over all core links in the network, if it is TE in the core that is of interest.

Note that worst-case utilisation is directly related to the *maximum resilient throughput* achievable in a network. This is the maximum throughput achievable, assuming that all demands grow uniformly, before utilisation in any link, in any failure scenario, exceeds a certain threshold. The maximum resilient throughput of a network is a useful number in the context of network design and planning, while worst-case utilisation is specifically of interest to traffic engineers, once a network topology is fixed.

Minimization of worst-case utilisation in a network is one simple goal that can guide traffic engineering optimisation procedures, as well as compare the performance of different TE methods on a particular network. It is often possible, though, to decrease worst-case utilisation in a network substantially by allowing routes to take paths far away from the shortest path to the destination; this may result in an unacceptable increase in latency. Hence, in practice, utilisation reduction must be accompanied by maintaining bounds on latency, or at least traded off against latency in some way.

Although we focus on worst-case utilisation as a goal in the study in section 5, other utilisation statistics could be used which are not determined entirely by the worst possible eventuality. Averages of the highest utilisation links, or a simple bound on worst-case utilisation while latency is minimized, are related criteria which could also be considered.

### 4.4 TRAFFIC DEMAND MATRIX

The network traffic demand matrix is a measure, estimation or prediction of the aggregated, e.g. router-to-router traffic flows across a network. By understanding the network traffic matrix and the network routing model it is possible to understand the impact on the network caused by changes to the network routing scheme, e.g. with different approaches to traffic engineering.

Demand level measurements have traditionally not been available in pure IP networks. A number of options, however, are now available to obtain this information:

- *IP flow statistics aggregation.* If edge devices are capable of accounting at a flow level (i.e. in terms of packet and byte counts), then a number of potential criteria could be used to aggregate this flow information – potentially locally on the device – in order to produce a traffic matrix. This information can be exported from routers using the proprietary Cisco Systems Netflow protocol [5], or the similar IP Flow Information eXport (IPFIX) protocol [RFC5101] which has been defined within the IETF.
- *MPLS LSP accounting.* Where MPLS is deployed an LSP implicitly represents an aggregate demand of traffic from the source of the LSP to the destination. Hence, if traffic accounting statistics are maintained per LSP, these can be retrieved to produce the network traffic matrix, using SNMP for example.
- *Demand estimation.* Demand estimation is the application of mathematical methods to measurements taken from the network, such as link usage statistics, in order to infer the traffic demand matrix that generated those usage statistics. Sophisticated demand estimation algorithms have been developed in the last few years – see for example, the approach taken in [6] – and there are a number of commercially available tools that use these, or similar, techniques in order to derive the traffic demand matrix. Section 4.3 describes how the worst-case utilisation may be calculated for a given network topology, a given set of metrics, and a given demand matrix. An estimate of this worst-case utilisation can be obtained by substituting an estimate of the demand matrix. Indeed, aggregate measures such as this utilisation statistic tend to be estimated very accurately even if the estimates of the individual demands themselves are subject to error.

Further details on the options for deriving a traffic demand matrix are provided in [7].

# 5. AN EMPIRICAL COMPARISON OF TRAFFIC ENGINEERING APPROACHES

## 5.1 FORMAT OF THE STUDY

We studied six networks; these comprised both Tier 1 and Tier 2 service provider networks, and a content-delivery network. Some of these were from the US, some from Europe, and some global. Five of these networks are currently operational, and one was still just a proposal at the time of the study.

The traffic demand matrices for the respective networks were obtained in a variety of ways. Three of the networks had meshes of MPLS-TE tunnels already deployed, from which the demands could be read directly. Demand deduction, as described in 4.4, was used to estimate the matrices of two more. For the proposed network, estimates were obtained of the market sizes in its various points of presence, and a simple gravity model was used to obtain a demand matrix from these estimates.

The statistic used to compare the performance of TE methods on these six networks was the worst-case failure utilisation, described in Section 4.3. The failure sets considered in all cases were the sets of all single circuit failures in the core of the networks. The worst-case failure utilisation was calculated for the following routing scenarios:

1. *Delay-based IGP metrics.* As an example of typical IP routing with unoptimised metrics, no MPLS tunnels are used, routing is pure IP with IGP link metrics chosen to be proportional to the delays, or latencies, of the circuits containing the links. The practicalities of taking such an approach have been discussed previously; see, for example, [8].
2. *Optimised IGP metrics.* No MPLS tunnels are used. Routing is pure IP, with IGP metrics chosen to minimize worst-case failure utilisation in the network.
3. *Optimised MPLS-TE explicit path options.* Primary and Secondary explicit MPLS-TE paths were chosen for each source/destination router-pair. The tunnel paths are computed offline to optimise worst-case failure utilisation.
4. *The best possible routing.* The best possible worst-case failure utilisation that can be obtained, routing the given demands in the network. This figure is calculated by taking each failure scenario in turn, and calculating the best possible routing for that failure scenario, independent of all others. That best routing is the multicommodity-flow network routing solution (see [1]). This solution allows complete flexibility in routing of all demands between all sources and destinations: any route may be taken for each demand, and the routes may be split in arbitrary proportions between different paths between the source and destination. This form of routing is clearly not achievable under any existing IP or MPLS routing schemes, but defines a bound on the best possible case that could be achieved by an approach in theory.

We have not included results for MPLS-TE dynamic path options, because in this case the tunnel routing will be different for different vendor's routers – e.g. as different vendor's implementations take different choices in tie-breaker situations where more than one possible path are available – and actual tunnel path information was not available for the networks considered. [5], however, shows that MPLS-TE with dynamic path options may result in a lower (i.e. better) worst-case failure utilisation than default IGP metrics, but this approach underperforms compared to what can be achieved with both MPLS-TE with explicit path options and IGP metric-based TE.

## 5.2 RESULTS

The results of the study are shown in Figure 2. For each of the six networks, the worst-case utilisation for scenarios 1-3 above is shown, *as a percentage of scenario 4*, the best possible utilisation level achievable.

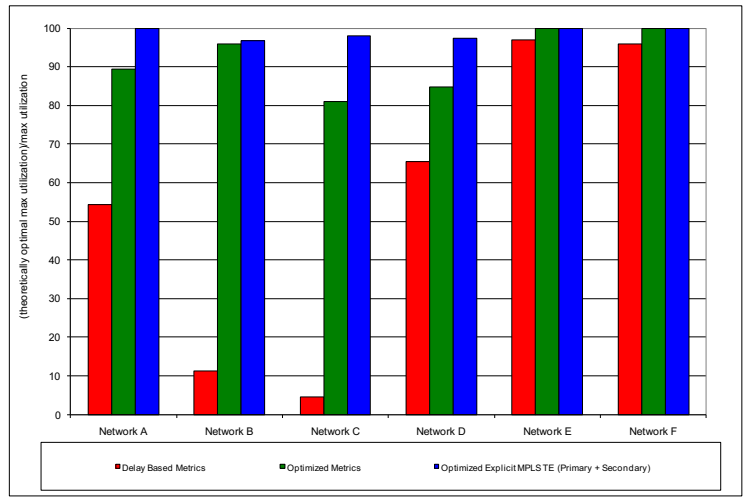

**Figure 2:** Results of Empirical Study of Relative TE Performances

## 5.3 CONCLUSIONS

The most noticeable feature of the results is the huge differences in the performance of delay-based metric routing across the six networks, ranging from less than 5% to over 95% of achievable utilisation. The networks for which delay-based routing did almost optimally were simpler, ring-based networks for which traffic engineering in general can achieve very little. In a typical ring-based topology, there are two paths from each source to each destination. If a circuit in the one path fails, the other path is taken, hence there is no scope for more sophisticated traffic engineering decisions. Also, in the networks in which delay-based metric routing performed extremely badly, performance could have been improved substantially by some obvious manual metric changes. However, all delay-based results are included here to provide a consistent, unoptimised performance measure.

Next, it is clear that MPLS-TE with explicit path options does outperform IGP metric-based routing in each of the six cases. What is perhaps surprising, though, is that the difference is not that large. MPLS TE with explicit path options achieves over 90% of the best possible utilisation levels, and metric based routing achieves over 80% of these levels, consistently across all six networks.

Under what circumstances does explicit routing do particularly well against metric-based routing? Very broadly, it seems that large networks with very heterogeneous circuit capacities seem to favour explicit routing more. To take a simple example, if a network contains a number of parallel circuits between two routers, and these circuits are of differing capacities, ECMP cannot split traffic adequately between these circuits. It can only split traffic evenly across circuits in this configuration, or omit some circuits entirely. This idea seems to extent to more complex situations of mismatched demands. These types of heterogenous networks tend to arise from the merger of two or more networks, rather than a single coherent network design. The differences still tend not to be too great, however, and it could be argued that in these situations some simple rationalizations of the network topologies themselves would produce utilisation gains out of proportion to these small percentage differences.